\documentclass[twocolumn,english,aps,prl,showpacs,amsmath,amssymb,superscriptaddress]{revtex4}
\usepackage[latin9]{inputenc}
\usepackage{color}
\usepackage{amstext}
\usepackage{amssymb}
\usepackage{graphicx}

\makeatletter
\@ifundefined{textcolor}{}
{%
 \definecolor{BLACK}{gray}{0}
 \definecolor{WHITE}{gray}{1}
 \definecolor{RED}{rgb}{1,0,0}
 \definecolor{GREEN}{rgb}{0,1,0}
 \definecolor{BLUE}{rgb}{0,0,1}
 \definecolor{CYAN}{cmyk}{1,0,0,0}
 \definecolor{MAGENTA}{cmyk}{0,1,0,0}
 \definecolor{YELLOW}{cmyk}{0,0,1,0}
}

\usepackage{color}

\newcommand{\beq}{\begin{eqnarray}}
\newcommand{\eeq}{\end{eqnarray}}

\makeatother
\usepackage{babel}

\begin{document}

\title{Nodeless superconductivity in the kagome metal CsV$_3$Sb$_5$}

\author{Weiyin Duan}
\affiliation{Center for Correlated Matter and Department of Physics, Zhejiang University, Hangzhou 310058, China}
\affiliation  {Zhejiang Province Key Laboratory of Quantum Technology and Device, Department of Physics, Zhejiang University, Hangzhou 310058, China}

\author{Zhiyong Nie}
\affiliation{Center for Correlated Matter and Department of Physics, Zhejiang University, Hangzhou 310058, China}
\affiliation  {Zhejiang Province Key Laboratory of Quantum Technology and Device, Department of Physics, Zhejiang University, Hangzhou 310058, China}

\author{Shuaishuai Luo}
\affiliation{Center for Correlated Matter and Department of Physics, Zhejiang University, Hangzhou 310058, China}
\affiliation  {Zhejiang Province Key Laboratory of Quantum Technology and Device, Department of Physics, Zhejiang University, Hangzhou 310058, China}

\author{Fanghang Yu}
\affiliation{Hefei National Laboratory for Physical Sciences at Microscale and Department of Physics, and CAS Key Laboratory of Strongly-coupled Quantum Matter Physics, University of Science and Technology of China, Hefei, Anhui 230026, China}

\author{Brenden R. Ortiz}
\affiliation{Materials Department and California Nanosystems Institute, University of California Santa Barbara, Santa Barbara, CA, 93106, United States}

\author{Lichang Yin}
\affiliation{Center for Correlated Matter and Department of Physics, Zhejiang University, Hangzhou 310058, China}
\affiliation  {Zhejiang Province Key Laboratory of Quantum Technology and Device, Department of Physics, Zhejiang University, Hangzhou 310058, China}
\author{Hang Su}
\affiliation{Center for Correlated Matter and Department of Physics, Zhejiang University, Hangzhou 310058, China}
\affiliation  {Zhejiang Province Key Laboratory of Quantum Technology and Device, Department of Physics, Zhejiang University, Hangzhou 310058, China}
\author{Feng Du}
\affiliation{Center for Correlated Matter and Department of Physics, Zhejiang University, Hangzhou 310058, China}
\affiliation  {Zhejiang Province Key Laboratory of Quantum Technology and Device, Department of Physics, Zhejiang University, Hangzhou 310058, China}
\author{An Wang}
\affiliation{Center for Correlated Matter and Department of Physics, Zhejiang University, Hangzhou 310058, China}
\affiliation  {Zhejiang Province Key Laboratory of Quantum Technology and Device, Department of Physics, Zhejiang University, Hangzhou 310058, China}

\author{Ye Chen}
\affiliation{Center for Correlated Matter and Department of Physics, Zhejiang University, Hangzhou 310058, China}
\affiliation  {Zhejiang Province Key Laboratory of Quantum Technology and Device, Department of Physics, Zhejiang University, Hangzhou 310058, China}

\author{Xin Lu}

\affiliation{Center for Correlated Matter and Department of Physics, Zhejiang University, Hangzhou 310058, China}
\affiliation  {Zhejiang Province Key Laboratory of Quantum Technology and Device, Department of Physics, Zhejiang University, Hangzhou 310058, China}

\author{Jianjun Ying}
\affiliation{Hefei National Laboratory for Physical Sciences at Microscale and Department of Physics, and CAS Key Laboratory of Strongly-coupled Quantum Matter Physics, University of Science and Technology of China, Hefei, Anhui 230026, China}

\author{Stephen D. Wilson}
\affiliation{Materials Department and California Nanosystems Institute, University of California Santa Barbara, Santa Barbara, CA, 93106, United States}

\author{Xianhui Chen}
\affiliation{Hefei National Laboratory for Physical Sciences at Microscale and Department of Physics, and CAS Key Laboratory of Strongly-coupled Quantum Matter Physics, University of Science and Technology of China, Hefei, Anhui 230026, China}
\affiliation{CAS Center for Excellence in Quantum Information and Quantum Physics, Hefei, Anhui 230026, China}
\affiliation{Collaborative Innovation Center of Advanced Microstructures, Nanjing 210093, People's Republic of China}

\author{Yu Song}
\email{yusong_phys@zju.edu.cn}
\affiliation{Center for Correlated Matter and Department of Physics, Zhejiang University, Hangzhou 310058, China}
\affiliation  {Zhejiang Province Key Laboratory of Quantum Technology and Device, Department of Physics, Zhejiang University, Hangzhou 310058, China}

\author{Huiqiu Yuan}

\email{hqyuan@zju.edu.cn}

\selectlanguage{english}%

\affiliation{Center for Correlated Matter and Department of Physics, Zhejiang University, Hangzhou 310058, China}
\affiliation  {Zhejiang Province Key Laboratory of Quantum Technology and Device, Department of Physics, Zhejiang University, Hangzhou 310058, China}
\affiliation{Collaborative Innovation Center of Advanced Microstructures, Nanjing 210093, People's Republic of China}
\affiliation  {State Key Laboratory of Silicon Materials, Zhejiang University, Hangzhou 310058, China}

\date{\today}

\begin{abstract}

The recently discovered kagome metal series $A$V$_3$Sb$_5$ ($A$=K, Rb, Cs) exhibits topologically nontrivial band structures, chiral charge order and superconductivity, presenting a unique platform for realizing exotic electronic states. The nature of the superconducting state and the corresponding pairing symmetry are key questions that demand experimental clarification. Here, using a technique based on the tunneling diode oscillator, the magnetic penetration depth $\Delta\lambda(T)$ of CsV$_3$Sb$_5$ was measured down to 0.07~K. A clear exponential behavior in $\Delta\lambda(T)$ with marked deviations from a $T$ or $T^2$ temperature dependence is observed at low temperatures, indicating a deficiency of nodal quasiparticles. Temperature dependence of the superfluid density and electronic specific heat can be described by two-gap $s$-wave superconductivity, consistent with the presence of multiple Fermi surfaces in CsV$_3$Sb$_5$. These results evidence nodeless superconductivity in CsV$_3$Sb$_5$ under ambient pressure, and constrain the allowed pairing symmetry.

\end{abstract}


\maketitle

The unique geometry of the kagome lattice leads to magnetic frustration \cite{Syozi1951,Han2012,Broholm2020}, topologically nontrivial electronic structures \cite{Ye2018,Liu2018,Kang2019}, and both electronic \cite{Kang2019,Lin2018,Yin2019} and magnon \cite{Chisnell2015} flat bands. Superconductivity with exotic pairing symmetries and properties were also predicted for the kagome lattice \cite{Ko2009,Yu2012,Wang2013,Kiesel2013,Mazin2014}, although physical realizations of such exotic superconductors have been limited. 

The recent discovery of superconductivity in the two-dimensional kagome metal series $A$V$_3$Sb$_5$ ($A$ = K, Rb, Cs) provides a much-desired platform to investigate potentially exotic superconducting states on the kagome lattice \cite{Ortiz2019,Ortiz2020,ortiz2020superconductivity,yin2021superconductivity}. In addition to superconductivity, these systems also exhibit a chiral charge order with unusual characteristics \cite{jiang2020discovery,zhao2021cascade,chen2021roton,li2021observation,tan2021charge,feng2021chiral}, topological band crossings \cite{Ortiz2019,Ortiz2020}, and a giant anomalous Hall effect in the absence of magnetic local moments \cite{Yang2020,Ortiz2020,kenney2020absence,yu2021concurrence}. The nature of the superconducting state in the presence of these electronic states, remains to be clarified. Nonetheless, tantalizing evidence for an unusual superconducting state has emerged from the observation of spin-triplet supercurrent in K$_{1-x}$V$_3$Sb$_5$ Josephson junctions \cite{wang2020proximityinduced} and possible Majorana bound states inside the superconducting vortex cores of CsV$_3$Sb$_5$ \cite{liang2021threedimensional}.

A hallmark of unconventional superconductivity is a sign-changing superconducting order parameter \cite{Scalapino2012,Stewart2017}. While direct phase-sensitive evidence for such a sign-change is difficult to obtain \cite{VanHarlingen1995,Tsuei2000}, for a number of pairing symmetries the sign-change mandates nodes in the superconducting order parameter (such as $d_{x^2-y^2}$-pairing in the cuprates), and can be probed through various experimental techniques. Thermal conductivity measurements on CsV$_3$Sb$_5$ down to 0.15~K indicated possible nodal quasiparticles \cite{zhao2021nodal}, 
and superconducting domes were found in the temperature-pressure phase diagrams of $A$V$_3$Sb$_5$ \cite{zhao2021nodal,chen2021double,du2021pressuretuned}, reminiscent of behaviors in established unconventional superconductors. On the other hand, the tunneling spectrum of CsV$_3$Sb$_5$ measured by scanning tunneling microscopy could be described by an anisotropic $s$-wave gap without nodes \cite{liang2021threedimensional}. To further clarify the pairing symmetry in the $A$V$_3$Sb$_5$ series, experiments sensitive to low-energy excitations are badly needed. 
 

In this work, the magnetic penetration depth and specific heat of the kagome metal CsV$_3$Sb$_5$ were measured to probe its superconducting gap structure. The change of the magnetic penetration depth reveals a clear exponential behavior down to 0.07~K, and deviates significantly from a $T$ or $T^2$ temperature dependence. Such a behavior suggests an absence of nodal quasiparticles, and instead points to fully-gapped superconductivity in CsV$_3$Sb$_5$ under ambient pressure. By analyzing the derived superfluid density and electronic specific heat, a two-gap $s$-wave superconducting order parameter
is found to
capture the experimental data. 
These results provide evidence for nodeless superconductivity in CsV$_3$Sb$_5$ under ambient pressure, and rule out pairing with symmetry-enforced nodes.

\begin{figure}
	\includegraphics[angle=0,width=0.48\textwidth]{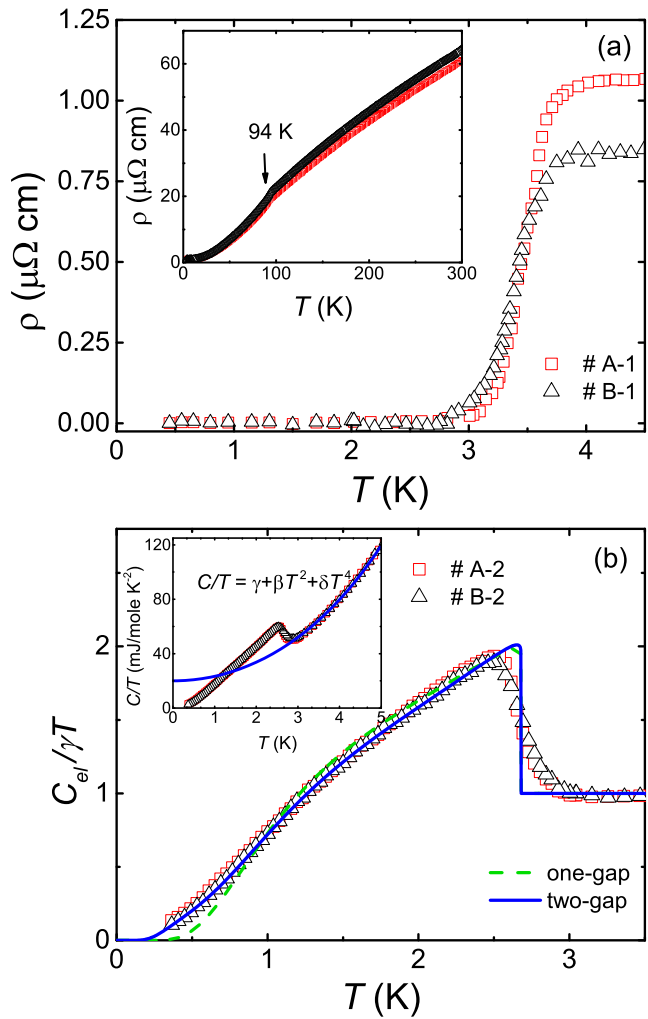}
	\vspace{-12pt} \caption{\label{figure1}(Color online) Temperature dependence of the low temperature (a) resistivity $\rho(T)$ and (b) electronic specific heat $C_{\rm el}(T)/\gamma T$ for CsV$_3$Sb$_5$. The inset in (a) shows $\rho(T)$ from 0.45~K up to 300~K. The inset in (b) shows the specific heat $C(T)/T$, with the blue solid line being a fit to the normal state specific heat, including contributions from electrons and phonons. The solid and dashed lines in (b) are fits to a single-gap and two-gap $s$-wave superconducting model, respectively.}
	\vspace{-12pt}
	\label{figure1}
\end{figure}

Single crystals of CsV$_3$Sb$_5$ were synthesized using the self-flux method, at the University of Science and Technology of China (sample A) and the University of California, Santa Barbara (sample B), with physical properties separately characterized and  described in previous works \cite{yu2021concurrence,Ortiz2020}. Eight samples were examined with various approaches, and are labeled samples \#A-1 to \#A-5, and samples \#B-1 to \#B-3. The $ab$-plane electrical resistivity $\rho(T)$ was measured in a $^3$He cryostat, using a standard four-probe method. Specific heat was measured using a Quantum Design Physical Property Measurement System (PPMS) with a $^3$He insert, using a standard pulse relaxation method. The change in the London penetration depth $\Delta\lambda(T)=\lambda(T)-\lambda(0)$ was measured using a tunnel diode oscillator (TDO) \cite{TDOdevice,TDO2}, with an operating frequency of about 7~MHz. The TDO method can precisely measure the temperature dependence of the magnetic penetration depth change, offering a powerful probe of low-energy excitations in the superconducting state. TDO measurements down to 0.35~K and 0.07~K, with noise levels as low as 0.1~Hz and 0.5~Hz, were carried out in $^3$He and dilution refrigerators, respectively.
The ac field generated by the coil (20~mOe) is far below the lower critical field $H_{\rm c1}$ of CsV$_3$Sb$_5$, ensuring that the sample remains in the full Meissner state throughout the measurements. $\Delta\lambda(T)$ can be obtained from the frequency shift $\Delta f(T)$ of the TDO through $\Delta\lambda(T)=G\Delta f(T)$, where $G$ is a sample-dependent scale factor \cite{Gfactor}.

The inset of Fig.~\ref{figure1}(a) shows the electrical resistivity $\rho(T)$ of CsV$_3$Sb$_5$ from 300~K down to 0.45~K. A clear kink can be observed around 94~K, due to the onset of charge order \cite{Ortiz2019,Ortiz2020}. The temperature evolution of $\rho(T)$ for samples \#A-1 and \#B-1 are highly similar, with residual resistivity ratios (RRR) of $\approx57$ and $\approx74$, respectively. Fig.~\ref{figure1}(a) zooms into $\rho(T)$ below 4.5~K, where a clear superconducting transition can be observed, which onsets around $3.5$~K and zero resistance appears below $\approx2.7$~K in both samples. The residual resistivities just above the onset of superconductivity are also similar, close to 1~$\mu\Omega$~cm in both samples.
Measurements of the low temperature specific heat $C(T)/T$ are shown in the inset of Fig.~\ref{figure1}(b) for samples \#A-2 and \#B-2, demonstrating the appearance of bulk superconductivity below $T_{\rm c}\approx2.7$~K. In the normal state, the specific heat can be modeled using $C(T)/T=\gamma+\beta T^2+\delta T^4$, with $\gamma~=~20.03$~mJ~mole$^{-1}$~K$^{-2}$, $\beta~=~3.306$~mJ~mole$^{-1}$~K$^{-4}$ and $\delta~=~26.78$ $\mu $J~mole$^{-1}$~K$^{-6}$ (solid line in the inset of Fig.~\ref{figure1}(b)). Here $\gamma$ is the Sommerfeld coefficient, and the other two parameters characterize the contribution from phonons. After subtracting the phonon contribution, the electronic specific heat $C_{\rm el}(T)/\gamma T$ can be obtained, shown in Fig.~\ref{figure1}(b). Analysis of $C_{\rm el}(T)/\gamma T$ will be further discussed below. 
Combining the coherence length $\xi$ ($\approx26$~nm from $H_{\rm c2}=0.47$~T \cite{Ortiz2020,yu2021concurrence}), residual resistivity $\rho_0$, and the Sommerfeld coefficient $\gamma$, mean free paths of 680~nm and 830~nm are estimated \cite{Orlando1979,Yuan2006} for samples \#A-1 and \#B-1, respectively. These values are much larger than the coherence length, indicating that the samples are in the clean limit.
Our resistivity and specific heat results are consistent with previous reports \cite{yu2021concurrence,Ortiz2020}, and indicate samples A and B are similar and that both are of high quality. 

Fig.~\ref{figure2} shows the temperature dependence of the magnetic penetration depth for CsV$_3$Sb$_5$.   
Three samples were studied with applied field along the $c$-axis, with $G = 7.8$~\AA/Hz, 9.1~\AA/Hz and 9.8~\AA/Hz, for samples \#A-3, \#A-4 and \#B-3, respectively. Sample \#A-5 was studied with field perpendicular to the $c$-axis, with its $\Delta\lambda(T)$ scaled to that of sample \#A-3, due to difficulties in accurately determining the $G$ factor for a thin sample.
The inset in Fig.~\ref{figure2}(a) shows $\Delta\lambda(T)$ for samples \#A-3 and \#A-5 from 4.5~K down to 0.07~K, exhibiting clear reductions upon cooling due to superconductivity, consistent with resistivity and specific heat measurements in Fig.~\ref{figure1}. 
$\Delta\lambda(T)$ with applied field along and perpendicular to the $c$-axis are compared in Fig.~\ref{figure2}(a) for samples \#A-3 and \#A-5, revealing almost identical behaviors
, suggesting that the superconducting state is rather isotropic. 
Fig.~\ref{figure2}(b) compares the low temperature penetration depth $\Delta\lambda(T)$ for three CsV$_3$Sb$_5$ samples with field applied along the $c$-axis. The data for different samples almost overlap, confirming the superconducting properties of samples A and B are similar, and demonstrate that our results are reproducible and reflect the intrinsic behavior of CsV$_3$Sb$_5$.


For a nodal superconductor in the clean limit, it is expected that the magnetic penetration depth exhibits a power-law behavior in the low temperature limit, {\it i.e.}, $\Delta\lambda\sim T^n$, with $n=1$ and 2 respectively corresponding to line nodes and point nodes being present in the gap structure. From Fig.~\ref{figure2}(b), it can be seen that the experimentally measured $\Delta\lambda(T)$ obviously deviates from a $T$ or $T^2$ behavior, but is reasonably described by $\Delta\lambda(T)\sim T^{2.9}$ with a slight deviation at low temperatures. To further analyze the power law dependence of $\Delta\lambda(T)$, the experimental data is fit by $\Delta\lambda(T)\sim T^n$ from 0.07~K to various temperatures, with the best fit $n$ shown in the inset of Fig.~\ref{figure2}(b). It can be seen that the exponent $n\approx3$ appears for $T>0.2 T_{\rm c}$, and becomes significantly enhanced at lower temperatures. Such a power law behavior ($n\gtrsim3$), in particular the progressive increase of $n$ with decreasing temperature, is consistent with an exponential behavior, and suggests an absence of gap nodes in the superconducting state of CsV$_3$Sb$_5$.

To further analyze the magnetic penetration depth, $\Delta\lambda(T)$ is fit to an $s$-wave gap at low temperatures,
with
\begin{equation}
\Delta\lambda(T)\sim T^{-\frac{1}{2}}\textrm{exp}\left(-\frac{\Delta(0)}{k_{\rm B}T}\right),
\label{equation1}
\end{equation}
\noindent where $\Delta(0)$ is the gap value at zero temperature. It can be seen that such an $s$-wave model fits the experimental data well at low temperatures, providing strong evidence for nodeless superconductivity in CsV$_3$Sb$_5$. The derived small superconducting gap of
 $\Delta(0)$~=~0.59~$k_{\rm B}T_{\rm c}$ 
indicates that $\Delta\lambda(T)$ would only saturate at very low temperatures, as seen in Fig.~\ref{figure2}(b), and a multi-gap model is needed in order to describe $\Delta\lambda(T)$.

\begin{figure}
	\includegraphics[angle=0,width=0.48\textwidth]{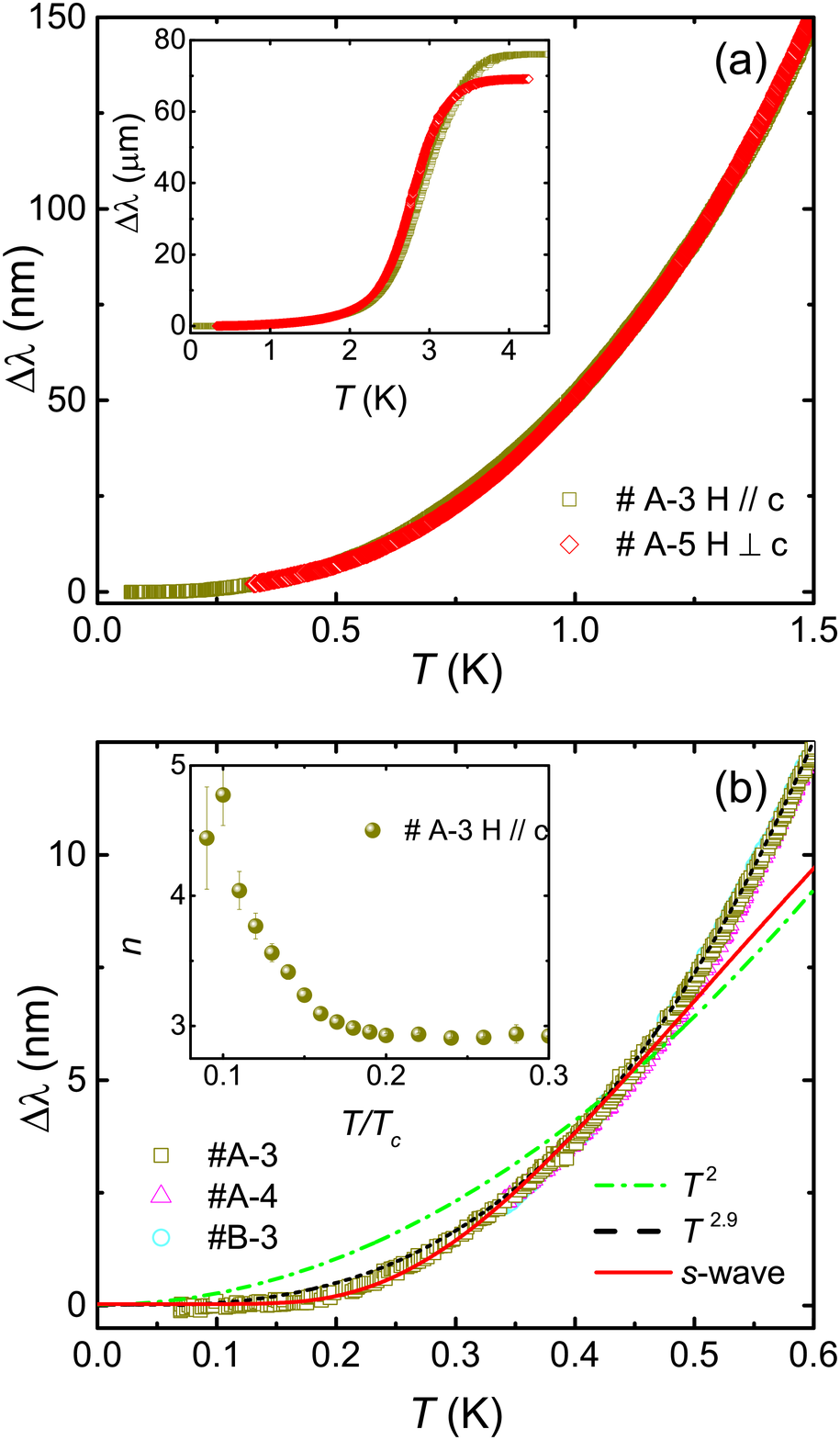}
	\vspace{-12pt} \caption{\label{figure2}(Color online) (a) Comparison of $\Delta\lambda(T)$ for applied fields along and perpendicular to the $c$-axis. The inset shows $\Delta \lambda(T)$ from 4.5~K down to 0.07~K. (b) The change of the penetration depth $\Delta\lambda(T)$, measured for three CsV$_3$Sb$_5$ samples with field along the $c$-axis. The dash-dotted and dashed lines respectively represent fits using a $T^2$ and a $T^n$ temperature dependence, from 0.07~K up to 0.6~K. The solid lines represents a fit to Eq.~\ref{equation1}. The inset shows the exponent $n$, obtained by fitting $\Delta\lambda(T)$ to a $T^n$ behavior from 0.07~K up to different temperatures of $T/T_{\rm c}$.}
	\vspace{-12pt}
	\label{figure2}
\end{figure}

To further extract information about the superconducting state, the normalized superfluid density is obtained through $\rho_s(T) = [\lambda(0)/\lambda(T)]^2$. The value of the zero-temperature penetration depth $\lambda(0)=387$~nm was estimated using $\lambda(0)=\sqrt{\phi_0H_{\rm c2}(0)}/\sqrt{24\gamma}\Delta(0)$ \cite{lamda0}, where $\phi_0$ is the magnetic-flux quantum, the Sommerfeld coefficient $\gamma$ is obtained from specific heat measurements (inset of Fig.~\ref{figure1}(b)), $H_{\rm c2}(0)=0.47$~T \cite{Ortiz2020,yu2021concurrence} and the weak-coupling limit of BCS theory with $\Delta(0)=1.76$~$k_{\rm B}T_{\rm c}$ is assumed. $\rho_{\rm s}$ obtained this way is shown as a function of the reduced temperature $T$/$T_{\rm c}$ for sample \#A-3 in Fig. \ref{figure3}(a), and is fit to several different models of the superconducting gap function $\Delta_k$, which is related to $\rho_{\rm s}$ through 

\begin{equation}
\rho_{\rm s}(T) = 1 + 2 \left\langle\int_{\Delta_k}^{\infty}\frac{E{\rm d}E}{\sqrt{E^2-\Delta_k^2}}\frac{\partial f}{\partial E}\right\rangle_{\rm FS},
\label{equation2}
\end{equation}

\noindent where $f(E,T)=[1+\exp(\frac{E}{k_{\rm B}T})]^{-1}$ is the Fermi-Dirac distribution, and $\langle\dots \rangle_{\rm FS}$ represents averaging over the Fermi surface. Temperature dependence of the superconducting order parameter is approximated using 

\begin{equation}
\Delta(T)~=~\Delta(0){\rm tanh}\left\{1.82\left[1.018\left(T_{\rm c}/T-1\right)\right]^{0.51}\right\}.
\label{equation3}
\end{equation}

\begin{figure}
	\includegraphics[angle=0,width=0.48\textwidth]{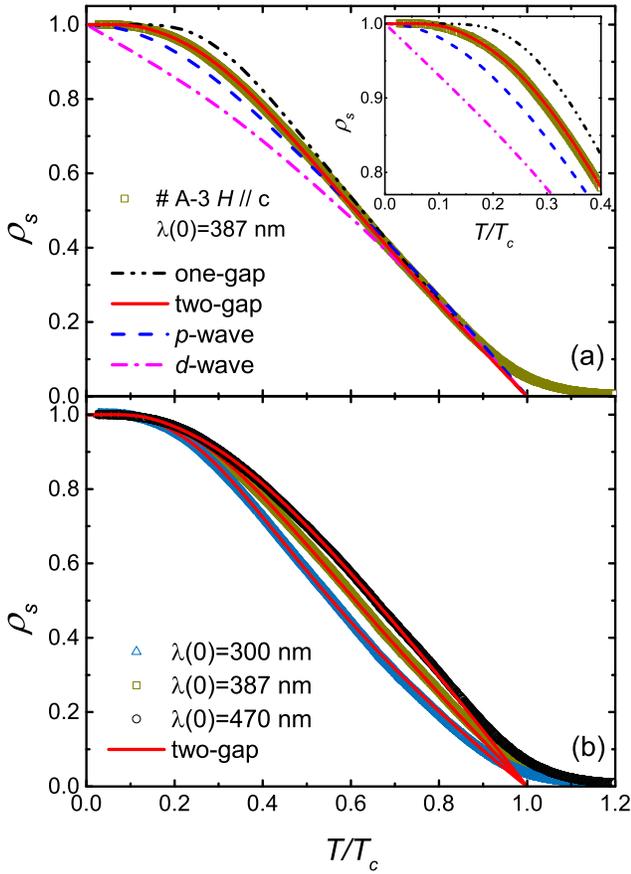}
	\vspace{-12pt} \caption{\label{figure3}(Color online) Normalized superfluid density $\rho_{\rm s}$ for sample \#A-3 with $\lambda(0)=387$~nm, as a function of the reduced temperature $T/T_{\rm c}$. The dash-dot-dotted, solid, dashed, and dash-dotted  lines respectively represent fits to models with a single $s$-wave gap, two $s$-wave gaps, a $p$-wave gap, and a $d$-wave gap. The inset zooms into the low temperature region of the figure. (b) Comparison of $\rho_{\rm s}$ for different values of $\lambda(0)$. The solid lines are fits to a two-gap $s$-wave model.}
	\vspace{-12pt}
	\label{figure3}
\end{figure}

As can be seen, the data are clearly incompatible with a single-gap $s$-wave model, a $p$-wave model with point nodes, and a $d$-wave model with line nodes. These models exhibit significant deviations from the data, especially at low temperatures (inset of Fig.~\ref{figure3}(a)).
On the other hand, a two-gap $s$-wave model describes the data well over the entire temperature range (solid line in Fig.~\ref{figure3}(a)).
The fit gap values are $\Delta_1(0)=1.42$~$k_{\rm B}T_{\rm c}$ and $\Delta_2(0)=0.58$~$k_{\rm B}T_{\rm c}$, with weights of the two gaps being 87\% and 13\%, respectively. The deduced small gap of $\Delta_2=0.58$~$k_{\rm B}T_{\rm c}$ is consistent with $\Delta = 0.59$~$k_{\rm B}T_{\rm c}$ derived from fitting $\Delta\lambda(T)$ at low-temperature (Fig.~\ref{figure2}(b)). 
Considering possible uncertainties in the estimate of $\lambda(0)$ and to test the robustness of our conclusions, $\lambda(0)$ is varied by $\approx\pm20$\% from 387~nm and the analyses are repeated, with results shown in Fig.~\ref{equation3}(b). It is found that the two-gap $s$-wave model describes $\rho_{\rm s}$ well in all cases (solid lines), while other models in Fig.~\ref{figure3}(a) exhibit clear deviations. For $\lambda(0)=300$~nm, $\Delta_1(0)=1.23$~$k_{\rm B}T_{\rm c}$ and $\Delta_2(0)=0.50$~$k_{\rm B}T_{\rm c}$ are obtained, while $\Delta_1(0)=1.63$~$k_{\rm B}T_{\rm c}$ and $\Delta_2(0)=0.66$~$k_{\rm B}T_{\rm c}$ are obtained for $\lambda(0)=470$~nm. Our analyses of the superfluid density therefore support the idea that CsV$_3$Sb$_5$ exhibits multiband nodeless superconductivity. It is noted that the 
larger gap $\Delta_1(0)$ is close to but still smaller than the BCS value of $1.76$~$k_{\rm B}T_{\rm c}$, the origin of which is not clear. It is possible that the contribution of an additional band or the effect of an anisotropic gap may lead to such a weak deviation. 

To confirm the above conclusion, the low temperature electronic specific heat $C_{\rm el}(T)/\gamma T$ of CsV$_3$Sb$_5$ is analyzed, shown in Fig.~\ref{figure1}(b).
Similar to the superfluid density, it is found that a single $s$-wave gap does not capture the behavior of $C_{\rm el}(T)/\gamma T$ (solid line in Fig.~\ref{figure1}(b)), while a two-gap $s$-wave model gives an excellent description of the experimental data over the full temperature range (dashed line in Fig.~\ref{figure1}(b)). The fit gap values are $\Delta_1(0)=1.62$~$k_{\rm B}T_{\rm c}$ and $\Delta_2(0)=0.63$~$k_{\rm B}T_{\rm c}$, with weights of 79\% and 21\%, respectively. These values are in reasonable agreement with analysis of the superfluid density, and supports the notion that CsV$_3$Sb$_5$ exhibits fully-gapped multiband superconductivity.


The low temperature exponential behavior of $\Delta\lambda(T)$ provides evidence for nodeless superconducitivty, while the extracted small gap and analysis of both the superfluid density and the electronic specific heat evidence multiband superconductivity. This is consistent with the presence of multiple Fermi surfaces revealed by angle-resolved photoemission spectroscopy measurements and electronic structure calculations \cite{Ortiz2020}. Such a picture is supported by the consideration that the pocket around $\Gamma$ and the Fermi surfaces around $M$ are mainly associated with Sb $p_z$ and V $d$ orbitals, respectively \cite{feng2021chiral}, and are likely associated with superconducting gaps of different sizes. It should be noted that our analyses do not exclude an anisotropic $s$-wave superconducting gap. However, considering that this is a multiband system, and the penetration depth along different directions are similar (Fig.~\ref{figure2}(a)), multi-gap superconductivity appears more likely.


Since a circular Fermi surface around $\Gamma$ is present in CsV$_3$Sb$_5$ \cite{Ortiz2020}, all even-parity basis gap functions associated with the $D_{6h}$ point group would lead to symmetry-enforced nodes, except that of the $A_{1g}$ channel \cite{Sigrist1991}. Therefore, the finding of nodeless superconductivity indicates the pairing symmetry of CsV$_3$Sb$_5$ belongs to the even-parity $A_{1g}$ representation, or odd-parity nodeless representations. While our results can be described by a superconducting state with two $s$-wave gaps, and is supported by the presence of multiple Fermis surfaces in CsV$_3$Sb$_5$, our measurements do not rule out some exotic unconventional superconducting states. These include spin-triplet nodeless superconductivity \cite{Balian1963}, $s^{\pm}$-pairing as in the iron-based superconductors \cite{Hirschfeld2011}, and pairing with band-mixing as suggested for CeCu$_2$Si$_2$ \cite{Pang2018}.

Upon pressure-tuning, superconductivity in CsV$_3$Sb$_5$ exhibits significant modulations and exhibit at least two superconducting domes \cite{zhao2021nodal,chen2021double}, suggesting the presence of competing superconducting ground states. 
As our results evidence a nodeless superconducting state under ambient pressure, whether the competing superconducting state stabilized under pressure remains nodeless or exhibits symmetry-enforced nodes, becomes an important question to be addressed in future works.

In conclusion, the superconducting pairing symmetry of CsV$_3$Sb$_5$ single crystals is probed through magnetic penetration depth measurements down to 0.07~K. A clear exponential behavior at low temperatures provides evidence for nodeless superconductivity in CsV$_3$Sb$_5$ under ambient pressure. Temperature dependence of the superfluid density and electronic specific heat can be described by two-gap $s$-wave superconductivity, consistent with the presence of multiple Fermi surfaces in CsV$_3$Sb$_5$. Our results are inconsistent with pairing with symmetry-enforced nodes in CsV$_3$Sb$_5$, but do not rule out fully-gapped unconventional superconductivity.


This work was supported by the National Key R\&D Program of China (No. 2017YFA0303100, No. 2016YFA0300202),the Key R\&D Program of Zhejiang Province, China (2021C01002), the National Natural Science Foundation of China (No. 11974306 and No. 12034017).  S.D.W. and B.R.O. gratefully acknowledge support via the UC Santa Barbara NSF Quantum Foundry funded via the Q-AMASE-i program under award DMR-1906325.  B.R.O. also acknowledges support from the California NanoSystems Institute through the Elings fellowship program.

\bibliography{bibfile}
\end{document}